\documentstyle[11pt,Moriond,epsfig]{article}

\def\be{\begin{equation}}
\def\ee{\end{equation}}
\def\bea{\begin{eqnarray}}
\def\eea{\end{eqnarray}}

\begin{document}
\vspace*{0mm}
\title{\boldmath FINDING $H^0 \rightarrow b\bar{b}$ AT THE LHC}

\author{ V. DROLLINGER }

\address{IEKP, University of Karlsruhe, Wolfgang-Gaede-Str. 1,\\
76131 Karlsruhe, Germany}

\maketitle\abstracts{
We have investigated the sensitivity of CMS for finding the Higgs boson in the $H^0 \rightarrow b\bar{b}$ channel. An excellent $b$-tagging performance and a good jet resolution are the main requirements needed for a successful event selection. In the Standard Model (SM), the $t\bar{t} H^0 \rightarrow l^\pm \nu q\bar{q} b\bar{b} b\bar{b}$ channel is accessible, if the Higgs mass is lighter than $m_{H^0} \approx$ 125 $GeV/c^2$, already during the first years of the LHC. Also, most of the MSSM (Minimal Supersymmetric Standard Model) parameter space can be covered. The $W^\pm H^0 \rightarrow l^\pm \nu b\bar{b}$ channel is only accessible with high luminosity at the LHC. In both channels the mass can be determined with a precision of a few per cent and the Higgs couplings at the level of 10\%.}

\section{General Situation}
Higgs bosons with a mass of around 115 ... 125 $GeV/c^2$ decay mainly to $b\bar{b}$ pairs. Therefore the $H^0 \rightarrow b\bar{b}$ channel \cite{DROLL} should contribute to a Higgs boson discovery~\footnote{If the Higgs boson is lighter than approximately 130 $GeV/c^2$.} at the LHC. However, the reconstruction of these events is experimentally challenging. Since the background which contains QCD jets is about two orders of magnitude higher than the signal, the $b$-tagging performance has to be very good. In CMS, the average $b$-tagging efficiency for $b$-jets coming from the Higgs decay is 60\%. At the same time, the mistagging probability for $c$-jets is 10\% and for light quark jets 1\%. This is enough to suppress the background sufficiently. In order to see a sharp Higgs mass peak on top of the non-resonant background, a good mass resolution is needed. Depending on the selection a resolution of 10\% to 15\% can be achieved. These numbers are comparable for the ATLAS and CMS experiments.

The CMS detector response is simulated with the GEANT3-based detector parametrisations CMSJET \cite{CMSJET} where the calorimeter response, jet-, lepton-, photon- and missing transverse energy reconstruction is simulated, and FATSIM \cite{FATSIM}. The latter is a parametrisation of momenta and impact parameters of tracks which is important to perform a realistic $b$-tagging simulation.

\section{$t\bar{t} H^0 \rightarrow l^\pm \nu q\bar{q} b\bar{b} b\bar{b}$}
\begin{figure}[ht]
\begin{center}
 \vspace*{-3mm}
 \includegraphics[width=0.60\textwidth,angle=+0]{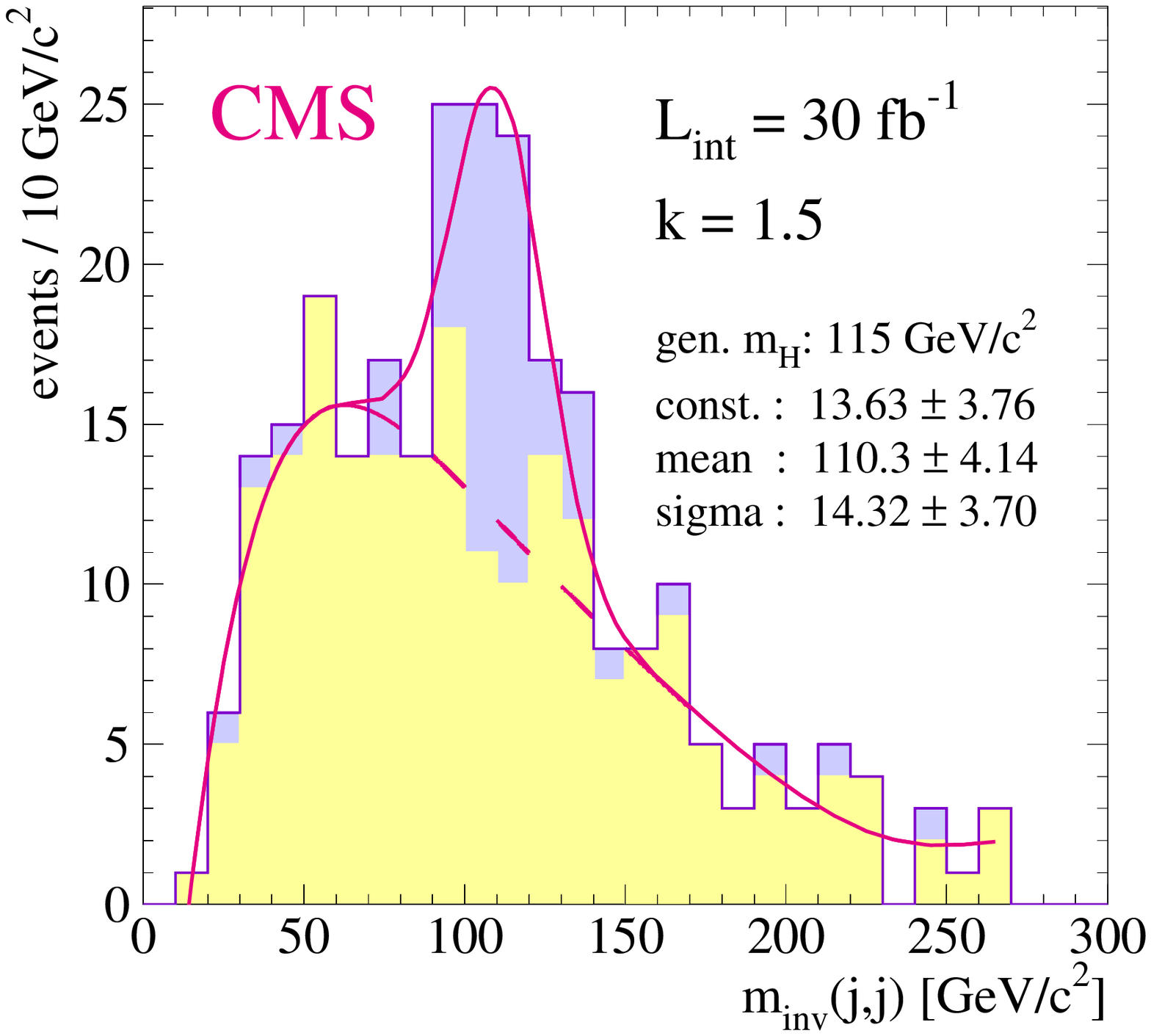}
 \includegraphics[width=0.39\textwidth,angle=+0]{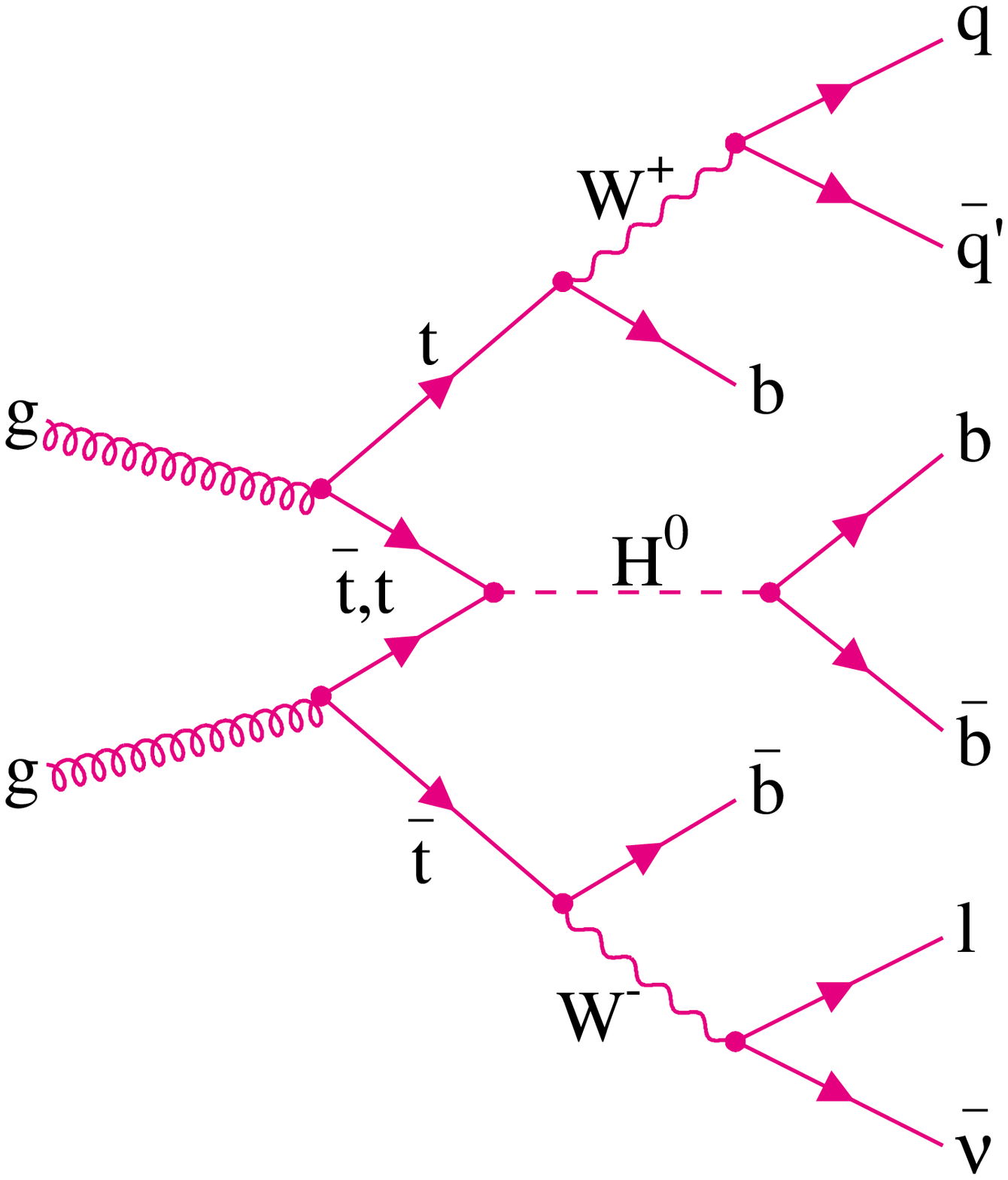}
 \caption{Left: simulated invariant mass distribution of signal (dark shaded, $m_{H^0} =$ 115 $GeV/c^2$) plus background for $L_{int} = $ 30 $fb^{-1}$. The dashed curve is obtained from the fit of the background without signal, the solid line describes the fit of signal plus background. Right: one example of $t\bar{t} H^0 \rightarrow l^\pm \nu q\bar{q} b\bar{b} b\bar{b}$ event at LO.
 \label{fig:tth_sm}}
 \vspace*{-6mm}
\end{center}
\end{figure}

All signal and background events of the $t\bar{t} H^0$ channel are generated with CompHEP \cite{CompHEP}, and the fragmentation is performed using the PYTHIA \cite{PYTHIA} program. The LO (leading order) cross sections at the LHC (proton proton collisions at $\sqrt{s} =$ 14 $TeV$) are summarised in Table~\ref{tab:tth_sm}. As visualised in Figure~\ref{fig:tth_sm} (right), the final state is expected to consist of one isolated lepton ($e^\pm$ or $\mu^\pm$), four $b$-jets, at least two additional jets and missing transverse energy from the neutrino.
\clearpage\newpage
$\ $\vspace*{-11mm}
\begin{table}[ht]
  \caption[LO Cross Sections for $t\bar{t} H^0$ Channel]{\sl CompHEP cross sections of SM signal and background relevant for the $t\bar{t} H^0 \rightarrow l^\pm \nu q\bar{q} b\bar{b} b\bar{b}$ channel and calculated with parton density function CTEQ4l \cite{PDFLIB}. The branching ratio of the semileptonic decay mode, not included here, is 29\% (only $W^\pm$ decays to  electrons or muons are taken into account) and $m_{W^\pm} =$ 80.34~$GeV/c^2$.\nolinebreak\rm}\label{tab:tth_sm}
\begin{center}
\begin{tabular}{|lcr|lcr|}
 \hline
 \multicolumn{3}{|c|}{LO cross sections} & \multicolumn{3}{|c|}{masses}\\
 \hline
  $\sigma_{t\bar{t}H^0} \times BR_{H^0 \rightarrow b\bar{b}}$ & = 
 & 1.09 - 0.32 $pb$ &  $m_{H^0}$ & = & 100 - 130 $GeV/c^2$ \\
    $\sigma_{t\bar{t}Z^0}$ & = & 0.65 $pb$ & $m_{Z^0}$ & = & 91.187 $GeV/c^2$\\
 $\sigma_{t\bar{t}b\bar{b}}$ & = & 3.28 $pb$ &   $m_{b}$ & = & 4.62 $GeV/c^2$\\
       $\sigma_{t\bar{t}jj}$ & = & 507  $pb$ &   $m_{t}$ & = &  175 $GeV/c^2$\\
 \hline
\end{tabular}
\end{center}
\end{table}

\subsection{Reconstruction}
Events are triggered and preselected if there is an isolated lepton ($p_T >$ 10 $GeV/c$ , $|\eta| <$ 2.4) and at least six jets ($E_T >$ 20 $GeV$ , $|\eta| <$ 2.5). In the next step, the best jet configuration in the event has to be found. This is done by defining an event likelihood function which takes into account the probabilities of jets to be $b$-jets or not, two reconstructed top masses, one~\footnote{Only the hadronically decaying $W^\pm$ can be reconstructed. The leptonic $W^\pm$ (nominal) mass is used to calculate the $z$-momentum of the neutrino which is needed for the top mass reconstruction.} reconstructed $W^\pm$ mass and the order of $b$-jet energies. After the best configuration is found which is the one with the highest value of the event likelihood function, the signal events are enriched cutting on values of three likelihood functions which take into account $b$-tagging, resonances and kinematics of the event. Finally, events in a mass window (90 ... 130 $GeV/c^2$) are counted.

\subsection{SM Higgs Results}
From the PYTHIA fragmentation we obtain an intrinsic k-factor of 1.9 for $t\bar{t}q\bar{q}$ background events. For $t\bar{t}H^0$ and  $t\bar{t}Z^0$ events a k-factor of 1.5 is used~\footnote{This estimated value is the same factor as evaluated for $t\bar{t}$ events.}. The Standard Model results with $m_{H^0} =$ 115 $GeV/c^2$ and $L_{int} = $ 30 $fb^{-1}$ are the following:
\vspace*{2mm}
\begin{center}
 \begin{tabular}{|ccc|}
 \hline
 $S =$ 38 &   $S / B =$ 73\%       & $\Delta y_t / y_t =$  13\% \\
 $B =$ 52 &   $S / \sqrt{B} =$ 5.3 &     $\Delta m / m =$ 3.8\% \\
 \hline
 \end{tabular}
\end{center}
\vspace*{2mm}
The mass precision is obtained from the Gaussian fit in Figure~\ref{fig:tth_sm} (left); the top Higgs Yukawa coupling $y_t$ is determined using the relation $y_t \sim \sqrt{N}$ and assuming a known coupling $y_b$. In the SM this channel can be discovered (5$\sigma$) for a Higgs mass up to 122 $GeV/c^2$ for $L_{int} = $~30~$fb^{-1}$.\nolinebreak

\subsection{MSSM Higgs Results}
\begin{figure}[ht]
\begin{center}
 \includegraphics[width=0.49\textwidth,angle=+0]{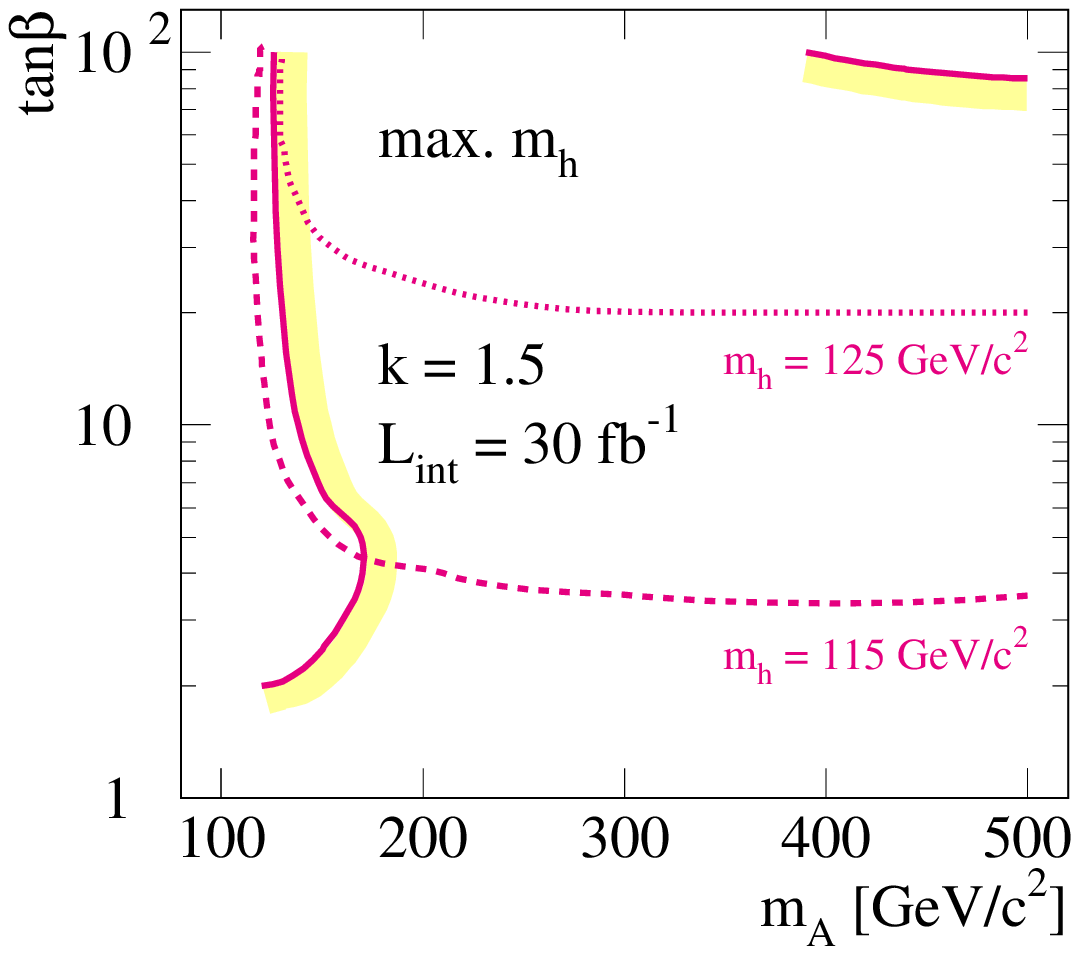}
 \includegraphics[width=0.49\textwidth,angle=+0]{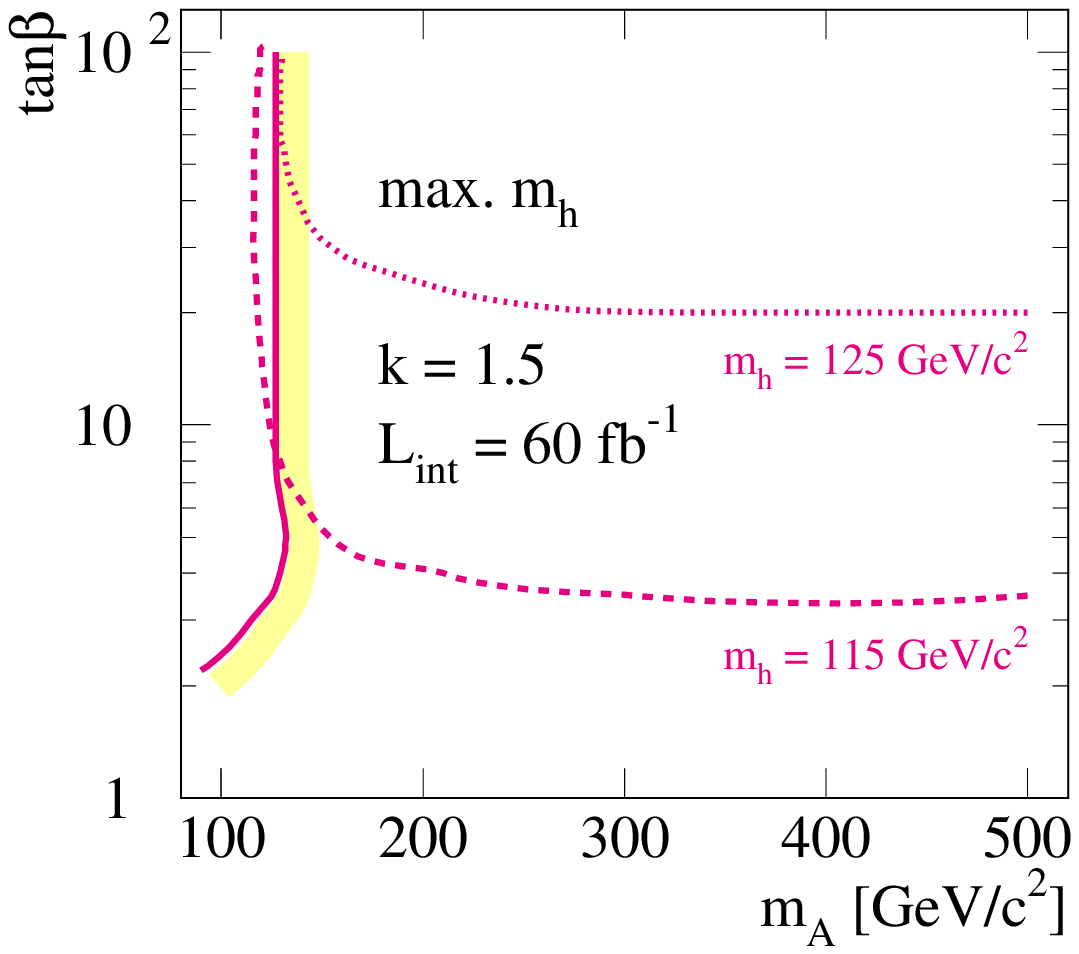}
 \caption{Discovery contours in the MSSM ("maximum $m_h$" scenario) parameter space for $L_{int} = $ 30 $fb^{-1}$ (left) and for $L_{int} = $ 60 $fb^{-1}$ (right). $S / \sqrt{B} \ge$ 5 to the shaded side of the solid line. The dotted and dashed lines are the isomass curves for $m_{h^0} =$ 125 $GeV/c^2$ and $m_{h^0} =$ 115 $GeV/c^2$, respectively.
 \label{fig:tth_susy}}
 \vspace*{-6mm}
\end{center}
\end{figure}

The SM results are used to calculate limits in the MSSM. Herefore the SM cross sections are compared with the MSSM cross section calculated with SPYTHIA \cite{SPYTHIA}. Figure~\ref{fig:tth_susy} shows the coverage of the MSSM parameter space in the worst scenario \cite{SUSYbenchmark} for two different integrated luminosities. Apart from the region at low $m_A$, the whole parameter space is covered in the $t\bar{t}h^0$ channel.

\section{$W^\pm H^0 \rightarrow l^\pm \nu b\bar{b}$}
Because it turned out that the $W^\pm H^0$ channel is only visible at high luminosity, the events are generated with PYTHIA including high luminosity pile up of minimum bias events. All important cross sections are listed in Table~\ref{tab:wh_sm}. In the final state, shown in Figure~\ref{fig:wh_sm} (right), we expect one isolated lepton, two $b$-jets and missing transverse energy.
\begin{table}[ht]
 \caption[LO Cross Sections for $W^\pm H^0$ Channel]{\sl PYTHIA \cite{PYTHIA} cross sections of signal (SM) and background, relevant for the $W^\pm H^0 \rightarrow l^\pm \nu b\bar{b}$ channel and calculated with parton density function CTEQ4l \cite{PDFLIB}. The cross sections do not include $BR_{W^\pm \rightarrow e^\pm \nu / \mu^\pm \nu} =$ 22\%.}\label{tab:wh_sm}
 \begin{center}
 \begin{tabular}{|lcr|lcr|}
 \hline
 \multicolumn{3}{|c|}{LO cross sections} & \multicolumn{3}{|c|}{masses from \cite{PDG98}}\\
 \hline
  $\sigma_{W^\pm H^0} \times BR_{H^0 \rightarrow b\bar{b}}$ & = 
 & 2.51 - 0.41 $pb$ &                  $m_{H^0}$ & = & 90 - 135 $GeV/c^2$ \\
 $\sigma_{W^\pm Z^0}$ & = & 27   $pb$ &   $m_{Z^0}$ & = & 91.187 $GeV/c^2$\\
  $\sigma_{W^\pm jj}$ & = & 30   $nb$ & $m_{W^\pm}$ & = & 80.41  $GeV/c^2$\\
  $\sigma_{t\bar{t}}$ & = & 570  $pb$ &     $m_{t}$ & = & 173.8  $GeV/c^2$\\
  $\sigma_{t\bar{b}}$ & = & 320  $pb$ &     $m_{b}$ & = & 4.3    $GeV/c^2$\\
 \hline
 \end{tabular}
 \end{center}
\end{table}

\subsection{Reconstruction}
Events are triggered by requiring one isolated lepton ($p_T >$ 20 $GeV/c$ , $|\eta| <$ 2.4) and two jets ($E_T >$ 30 $GeV$ , $|\eta| <$ 2.5). The event selection is based on a cut selection which includes $b$-tagging of two jets, veto on additional jets ($E_T >$ 20 $GeV$ , $|\eta| <$ 2.5) or leptons ($p_T >$ 5 $GeV/c$), transverse energy balance of the event, reconstruction of the transverse $W^\pm$ mass and a mass window around the Higgs mass peak (97 ... 130 $GeV/c^2$).

\subsection{SM Higgs Results}
\begin{figure}[ht]
\begin{center}
 \vspace*{-2mm}
 \includegraphics[width=0.60\textwidth,angle=+0]{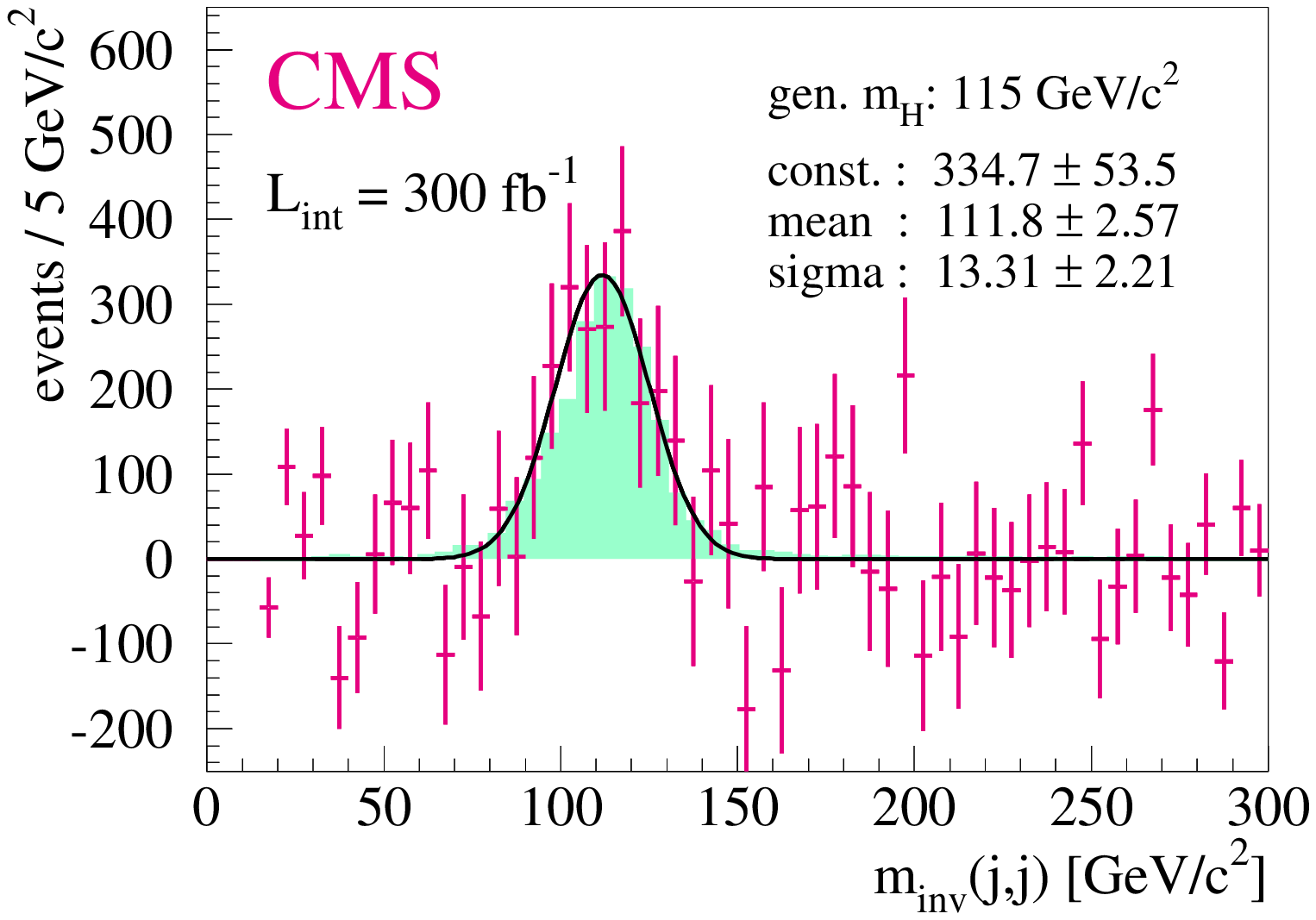}
 \includegraphics[width=0.39\textwidth,angle=+0]{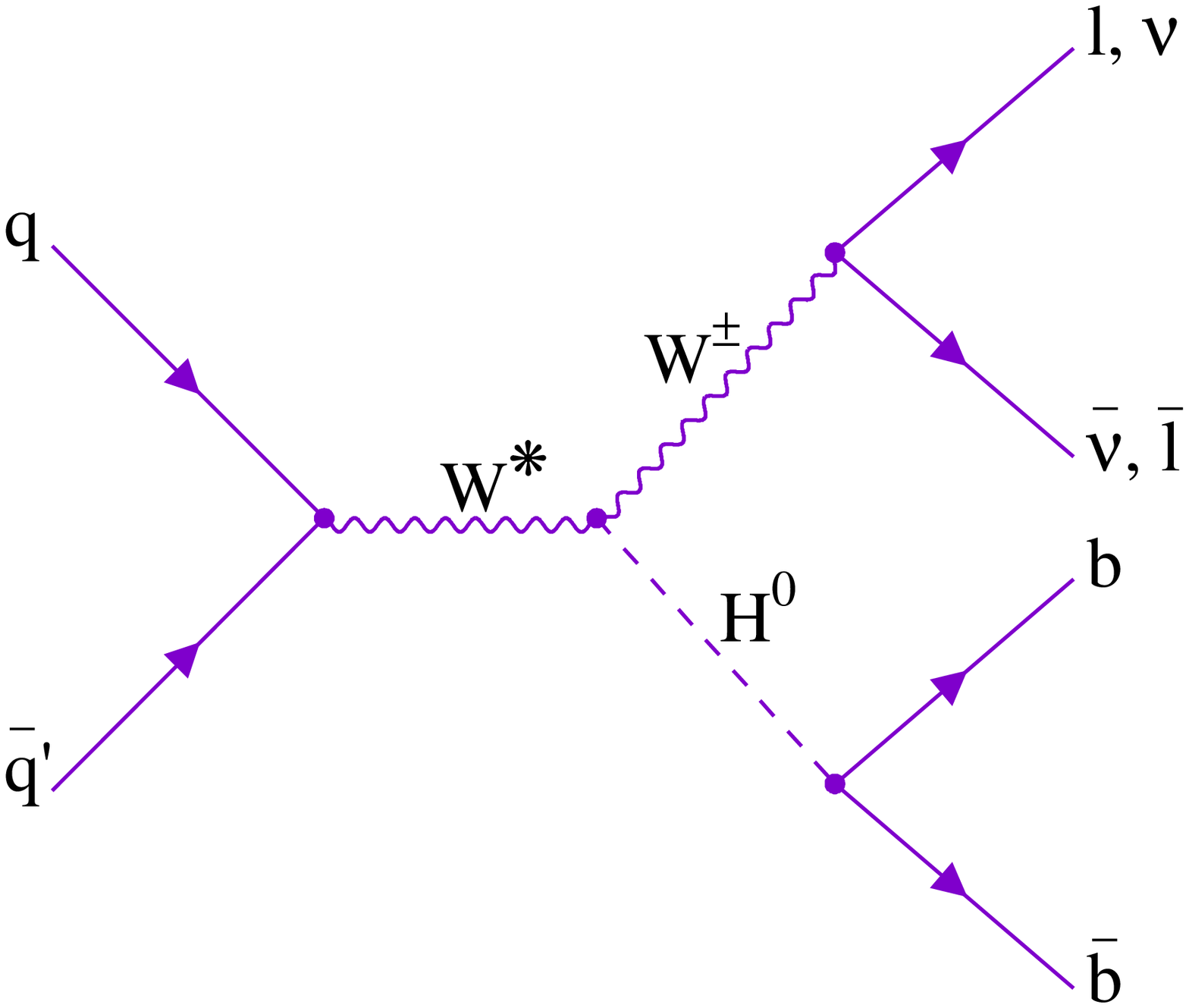}
 \vspace*{-5mm}
 \caption{Left: signal plus background after subtraction of all backgrounds. The simulated data points are fitted with a Gaussian which describes the pure $W^\pm H^0$ signal (shaded) well. Right: $W^\pm H^0 \rightarrow l^\pm \nu b\bar{b}$ event at LO.
 \label{fig:wh_sm}}
 \vspace*{-6mm}
\end{center}
\end{figure}

The SM results (without k-factors) for $m_{H^0} =$ 115 $GeV/c^2$ an $L_{int} = $ 300 $fb^{-1}$ are the following:\nolinebreak
\vspace*{2mm}
\begin{center}
 \begin{tabular}{|ccc|}
 \hline
 $S =$  1610 &  $S / B =$ 2.3\%       & $\Delta g_{WWH} / g_{WWH} =$ 8.4\% \\
 $B =$ 70948 &   $S / \sqrt{B} =$ 6.0 &             $\Delta m / m =$ 2.3\% \\
 \hline
 \end{tabular}
\end{center}
\vspace*{1mm}
The $WWH$ coupling is determined using the relation $g_{WWH} \sim \sqrt{N}$ and assuming a known coupling $y_b$; the mass precision is obtained from the Gaussian fit in Figure~\ref{fig:wh_sm} (left). For $L_{int} = $ 300 $fb^{-1}$ this channel is visible above 5$\sigma$ up to $m_{H^0} =$ 123 $GeV/c^2$ in the SM~\footnote{The MSSM results in this channel are qualitatively similar as in the $t\bar{t}h^0$ channel, but roughly ten times more luminosity is needed in the $W^\pm h^0$ channel.}.

\small{
\section*{Acknowledgments}
\vspace*{-1mm}
We would like to thank Daniel Denegri and Thomas M\"uller for an excellent support of this study and all the other people who participated in many fruitful discussions. Finally, many thanks to Viacheslav Ilyin who generated all CompHEP events for the $t\bar{t} H^0$ channel.

}

\end{document}